\documentclass{emulateapj}
\usepackage{amsmath}
\usepackage{amssymb}
\usepackage{graphicx}

\usepackage[usenames,dvips]{color}

\usepackage{natbib}
\newcommand\K{{\rm K}} 
\newcommand\s{{\rm s}} 
\newcommand\yr{{\rm yr}} 
\newcommand\Gyr{{\rm G}\yr} 
\newcommand\m{{\rm m}} 
\newcommand\mum{\mu\m} 
\newcommand\mm{{\rm m}\m} 
\newcommand\cm{{\rm c}\m} 
\newcommand\pc{{\rm pc}} 
\newcommand\kpc{{\rm k}\pc} 
\newcommand\au{{\rm AU}} 
\newcommand\erg{{\rm erg}} 
\newcommand\muas{\mu{\rm as}} 

\renewcommand\d{{\rm d}}

\def\Ms{{M_\odot}}

\begin{document}

\title{The Event Horizon of Sagittarius A*}

\author{
Avery E.~Broderick\altaffilmark{1},
Abraham Loeb\altaffilmark{2} \&
Ramesh Narayan\altaffilmark{2}
}
\altaffiltext{1}{Canadian Institute for Theoretical Astrophysics, 60 St.~George St., Toronto, ON M5S 3H8, Canada; aeb@cita.utoronto.ca}
\altaffiltext{2}{Institute for Theory and Computation, Harvard University, Center for Astrophysics, 60 Garden St., Cambridge, MA 02138.}

\shorttitle{The Event Horizon of Sgr A*}
\shortauthors{Broderick, Loeb \& Narayan}

\begin{abstract}
Black hole event horizons, causally separating the external universe
from compact regions of spacetime, are one of the most exotic
predictions of General Relativity (GR).  Until recently, their compact size
has prevented efforts to study them directly.  Here we show that recent
millimeter and infrared observations of Sagittarius A* (Sgr A*), the
supermassive black hole at the center of the Milky Way, all but
requires the existence of a horizon.  Specifically, we show that these
observations limit the luminosity of any putative visible compact
emitting region to below $0.4\%$ of Sgr A*'s accretion luminosity.
Equivalently, this requires the efficiency of converting the
gravitational binding energy liberated during accretion into radiation
and kinetic outflows to be greater than $99.6\%$, considerably larger
than those implicated in Sgr A*, and therefore inconsistent with the
existence of such a visible region.  Finally, since we are able to frame
this argument entirely in terms of observable quantities, our results
apply to all geometric theories of gravity that admit stationary
solutions, including the commonly discussed $f(R)$ class of theories.
\end{abstract}


\maketitle

\section{Introduction}

The Schwarzschild metric presents the first example of a compact
horizon: an imaginary surface delineating a compact region from which
the rest of the universe is causally disconnected.  Subsequently, a
substantial theoretical effort was made to determine if such ``black
hole'' solutions could practically come into existence
\citep{Oppe-Snyd:39,Penr:65,Whee:66,Hege_etal:03}.  Despite Einstein's
misgivings, black holes are now believed to be the inevitable
consequence of the demise of massive stars.  In recent years, the existence of compact
horizons has taken on a renewed significance due to
efforts to construct a Grand Unified Theory.  Specifically, horizons
play prominently in the well known ``information paradox''
\citep{Hawk:05,Math:08}.  However, now we are in a position to
constrain the existence of horizons observationally.

For astrophysical purposes, the critical features of horizons are (1)
their compactness, allowing substantial amounts of gravitational
binding energy to be liberated during accretion, and (2) their
ability to hide the ultimate fate of accreting matter.  Unlike a black
hole, into which the kinetic and thermal energy gained during infall
can disappear (adding only to the mass), objects with surfaces (e.g.,
stars) radiate the residual energy not emitted during the accretion
process. This fact has been used in a number of efforts to test for
the presence of horizons in a variety of black hole systems
\citep{Nara-Garc-McCl:97,Nara-Heyl:02,McCl-Nara-Rybi:04,Brod-Nara:06,Brod-Nara:07,Nara-McCl:08}.
For many black hole candidates the tell-tale sign of surface
emission, seen in accreting neutron stars, is absent, implying the
lack of an analogous surface in these objects.

Sagittarius A* (Sgr A*), the radio point-source associated with the
dark mass located at the center of the Milky Way, is the best studied
black hole candidate to date.  Near-infrared (NIR) observations of
massive stars in its vicinity have provided direct mass and distance
measurements, $M=4.5\pm0.4\times10^6\,\Ms$ and $D=8.4\pm0.4\,\kpc$,
respectively, and confined it to within $40\,\au$
\citep{Ghez_etal:08,Gill_etal:08}.  With a luminosity of 
$10^{36}\,\erg\,\s^{-1}$, it is substantially underluminous relative
to its limiting Eddington luminosity, $6\times 10^{44}\,\erg\,\s^{-1}$.
The emission is strongly non-thermal, distributed from the radio to
$\gamma$-rays, and believed to be powered by the release of gravitational
binding energy by accreting gas.

Due to its mass and proximity, Sgr A* has the largest angular size of
any known black hole and offers the best prospects for direct imaging
of its silhouette
\citep{Falc-Meli-Agol:00,Brod-Loeb:05,Brod-Loeb:06,Brod-Loeb:06b}.
Unfortunately, mm-imaging alone will be unable to verify or exclude
the presence of a horizon \citep{Brod-Nara:06}.  This is because even
a relatively bright central object (e.g., one radiating thermally with
a luminosity comparable to $\dot{M} c^2$) will appear in silhouette against
the much hotter surrounding accretion flow.  However, here we show
that recent mm-Very Long Baseline Interferometric (VLBI) observations,
which have resolved sub-horizon scale structure for the first time
\citep{Doel_etal:08}, provide
conclusive evidence for the presence of a horizon when coupled with
the existing NIR and mid-infrared (MIR) flux limits
\citep{Ghez_etal:05b,Horn_etal:07,Scho_etal:07}.  We do this by
explicitly excluding emission from a putative compact infrared
photosphere, lying inside of the mm-emitting region.  Specifically, we
show that for such surface emission to remain undetected would require
unphysically large radiative efficiencies; greater than $99.6\%$, as
compared to the typical efficiencies in AGN of $10\%$ \citep{Kato:08}
and the meager efficiency implicated in Sgr A*, $0.01$--$1\%$
\citep{Nara-McCl:08}.  Most importantly, we do this using only
observable quantities, ensuring that these limits are largely
independent of the gravitational theory employed\footnote{Our results
  hold for any geometric gravitational theory admitting stationary
  solutions.  This includes all of the $f(R)$ theories as well as
  black hole alternatives within the context of General Relativity,
  such as clusters of compact objects, boson stars and gravastars.}.

\S\ref{sec:PA} describes the underlying physical assumptions and
\S\ref{sec:OLUtEoH} presents the recent observational constraints,
within the context of Sgr A*.  Conclusions are collected in
\S\ref{sec:C}.  Appendices \ref{sec:GBEiSS} and \ref{sec:ASoCO}
discuss photon propagation times, the notion of gravitational binding
energy and the apparent size of objects within the context of general
geometric theories of gravity.

\section{Physical Assumptions}\label{sec:PA}
Our argument depends critically upon three underlying, physically
well motivated assumptions: (1) Sgr A* is accretion powered, (2) has
reached steady state and (3) compact surfaces are
approximate blackbodies.  We wish to emphasize that relaxing any
of these would require fundamental alterations to presently well
understood physics (such as microscopic physics at pedestrian, and
laboratory accessible, densities, temperatures and magnetic field
strengths).  Nevertheless, we discuss each in detail, describing how
these are justified in the context of Sgr A*.

\subsection{Accretion Power \& the Luminosity of Sgr A*}
It has been widely accepted that the emission from Sgr A* is
powered by accretion.  This is inferred from a variety of sources,
including the spectrum, variability, VLBI observations, environment
and polarization of Sgr A*.  Here we review these, if only to list the
numerous observational hurdles facing any alternate interpretation.

The spectrum of Sgr A* extends from the radio to the gamma-rays,
peaking in the sub-mm, where it transitions from a featureless,
inverted power-law to an optically thin spectrum.  Sgr A*'s emission
is strongly non-thermal and exhibits no absorption lines, which
despite it's low bolometric luminosity (comparable to a B-star) rules
out a hydrostatic, optically thick gas cloud, similar to a large star.
Furthermore, this spectrum is very similar to that from other AGN, for
which there is no known alternative power source with the necessary
efficiency.

At cm-wavelengths and below, Sgr A* exhibits considerable short
timescale variability.  Flares have been observed in the mm, sub-mm,
infrared and X-ray bands, with rise times and variability timescales
comparable to the periods of innermost stable orbits
around General Relativistic (GR) black holes.  At considerably longer
wavelengths, where Sgr A* is optically thick, the degree
of variability decreases and the characteristic timescales increase.
This is consistent with a stratified emitting region, with
short-wavelength emission arising close to the central mass.

This interpretation is confirmed by VLBI observations.  Observations
with the Very Long Baseline Array at wavelengths from $3\,\mm$ to
$6\,\cm$ have resolved the intrinsic size of the radio emitting region
about Sgr A* \citep{Bowe_etal:06,Shen_etal:05,Kric_etal:06}.  This is
somewhat complicated by interstellar scattering, requiring a careful
subtraction of the scattering law determined by fitting the
observed size of Sgr A* from $6\,\cm$ to $24\,\cm$.  While there is
some debate over the precise form of the scattering law, there is no
question that Sgr A* is indeed radially stratified, with the radius of the
photosphere decreasing with decreasing wavelength.

Turning to it's environment, Sgr A* suffers from an embarrassment of
riches.  The stellar winds from nearby massive stars
($\simeq0.1\,\pc$ from Sgr A*) provide sufficient material to support
luminosities more than 11 orders of magnitude greater than that
observed!  Indeed, the difficulty in modeling Sgr A* has been to
explain its meager luminosity instead of the prodigious output
predicted.  This has typically been done by postulating an
extraordinarily low radiative efficiency and/or the existence of
accretion-powered outflows \citep{Nara-Yi-Maha:95,Blan-Bege:99}.

Finally, below roughly $3\,\mm$, Sgr A* is linearly polarized
\citep{Aitk_etal:00,Bowe-Wrig-Falc-Back:03,Macq_etal:06,Marr-Mora-Zhao-Rao:06,Marr-Mora-Zhao-Rao:07}.
At longer wavelengths it is Faraday depolarized, providing some measurement of
the density and magnetic field strength near the central mass.
Assuming near equipartition magnetic fields, this implies a density
of cold electrons of roughly $10^6\,\cm^{-3}$ at radii of
$10^{13-14}\,\cm$ \citep{Agol:00,Quat-Gruz:00,Marr-Mora-Zhao-Rao:06}.  Assuming stronger, large-scale magnetic
fields results in lower density estimates, and conversely, large
amounts of magnetic turbulence results in higher estimates.  While
this does not provide a direct measurement of the accretion rate, it
is difficult to imagine that the magnetorotational instability is
incapable of driving accretion in this environment at rates up to
$10^{-8}\,\dot{\Ms}/\yr$.

%
%

\subsection{Steady State} \label{sec:PA.S}
Assuming steady state will allow us to relate the surface emission
from black hole alternatives to that from the accretion flow.  Within
the context of Sgr A* there are a variety of reasons to expect that
any black hole alternative will have reached some sort of steady
state \citep[see ][]{Brod-Nara:07}.
However, we point out that black holes explicitly violate this
condition: the unradiated kinetic energy is advected across the horizon
and then added to the mass of the black hole.  

The dynamical timescale of Sgr A* depends upon the nature of the
object.  Nevertheless, we may expect that it is comparable to the
dynamical timescale of the corresponding black hole,
$GM/c^3\simeq20\,\s$.  This is supported by the $\sim10\,\min$
variability observed in the NIR and X-rays, presumed to be associated
with material orbiting nearby.  Both of these are much, much shorter
than the estimated age of Sgr A*, $10\,\Gyr$, making it natural to
assume that it has sufficient time to reach steady state.  Note that
if a substantial portion of Sgr A*'s mass was accrued via accretion,
its accretion rate must have been much larger at times in the past.
Thus if the timescale for reaching steady state exceeds the period
since the previous active phase we will be {\em underestimating} Sgr
A*'s luminosity.

While the extreme difference in timescales naturally implies that even
surfaces with extraordinary redshifts ($z\lesssim 10^{15}$) will have
reached steady state, within the context of black hole alternatives
exhibiting GR exterior spacetimes a much stronger
statement can be made.  This is because photons that are emitted
initially outward and subsequently lensed back onto Sgr A* necessarily
provide a mechanism to couple otherwise disparate regions of the
photosphere.  The travel times of these photons diverge only
logarithmically with photosphere redshift, and therefore place an
upper bound upon the equilibration timescale.  

Inherent in these arguments is the assumption that {\em locally} the
surface properties of any black hole alternative are not ill-behaved.
That is, that the self-coupling is what determines the equilibration
timescale.  This may not be the case if, e.g., the heat capacity of
the surface is effectively nearly infinite (or as in the case of a
black hole, negative!).  However, if the rate at which accreted
baryonic material is incorporated into the surface is sufficiently
slow, a baryonic atmosphere will develop. In this case, even for
objects with near-vanishing temperature a hot atmosphere can be
produced which satisfies the steady state condition.

%

\subsection{Compact Objects \& Blackbodies}\label{sec:PA.COaB}
\begin{figure}
\begin{center}
\includegraphics[width=\columnwidth]{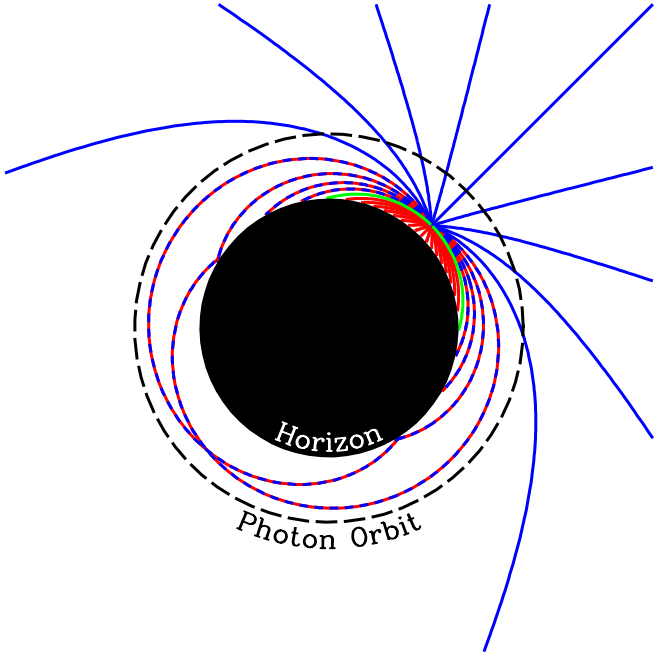}
\end{center}
\caption{Rays launched isotropically (every $10^\circ$) in the locally
  flat, stationary frame are lensed in a Schwarzschild spacetime.
  Those rays that are initially moving inwards, tangentially and
  outwards are shown in red, green and blue, respectively.
  Additionally, those that are launched initially moving
  outwards and are subsequently captured are red-blue dashed.  For
  reference the horizon and photon orbit are shown.  Generically, the
  fraction of rays that escape to infinity decreases as the emission
  point is moved towards the black hole, dropping below 50\% at the
  photon orbit and dropping all the way to 0\% at the horizon.  As a
  consequence of this strong lensing, emitting objects that are
  contained within the photon orbit approximate the canonical pin-hole
  cavity example of a blackbody, becoming a perfect blackbody in the
  limit that the surface redshift goes to $\infty$.
}\label{fig:pinhole}
\end{figure}

Our argument will hinge upon our ability to specify the spectral
signatures of any surface emission from black hole alternatives.
Subject to the condition outlined in \S\ref{sec:PA.S}, it is
natural to expect any such emission to be thermal.  This is
obvious if the object accrues an optically thick atmosphere of baryonic
material.  However, for any compact surface which lies within the
photon orbit this is generally an excellent assumption.

By definition, at the photon orbit whether or not a photon impacts the
surface is determined by the radial component of the momentum.  Thus,
if photons are emitted isotropically by a surface located at the
photon orbit, half will impact the surface and half will escape to
infinity.  As the surface shrinks inside the photon orbit the
fraction of escaping photons also decreases, falling to zero when the
surface coincides with the horizon.  Thus, the higher the
redshift of the surface, the more effectively different regions on the
surface are coupled to one another via radiation and the closer the
surface approaches thermodynamic equilibrium.  Consequently,
high-redshift surfaces present a perverse realization of the canonical
pin-hole cavity, becoming ideal blackbodies as $z$ goes to
$\infty$ \citep{Brod-Nara:06}.  For a Schwarzschild spacetime this is
shown in Fig. \ref{fig:pinhole}; however this behavior is generic to
spherically symmetric spacetimes.  Thus if the system has sufficient
time to have reached steady state, it must be a blackbody.

%

\section{Observational Limits Upon the Existence of Horizons}\label{sec:OLUtEoH}

The primary astrophysical importance of a horizon is that the
gravitational binding energy liberated by material as it accretes can
be advected into the black hole without any further observational
consequence.  This is very different from accretion onto other compact
objects, e.g., neutron stars, in which this liberated energy
ultimately must be emitted by the stellar surface.  Importantly,
this argument is not dependent upon the particulars of the compact
object.  Any object powered by accretion, whose surface is visible
from the external universe, should show evidence of surface
radiation.  We will use this fact to rule out the possibility that
accreted material in Sgr A* settles in a region visible to
outside observers, and in doing so make the argument that a horizon
{\em must} exist.  That is, we imagine that material comes to rest at
some surface where it radiates its remaining kinetic energy and
observationally constrain the associated surface luminosity.

\begin{figure}
\begin{center}
\includegraphics[width=\columnwidth]{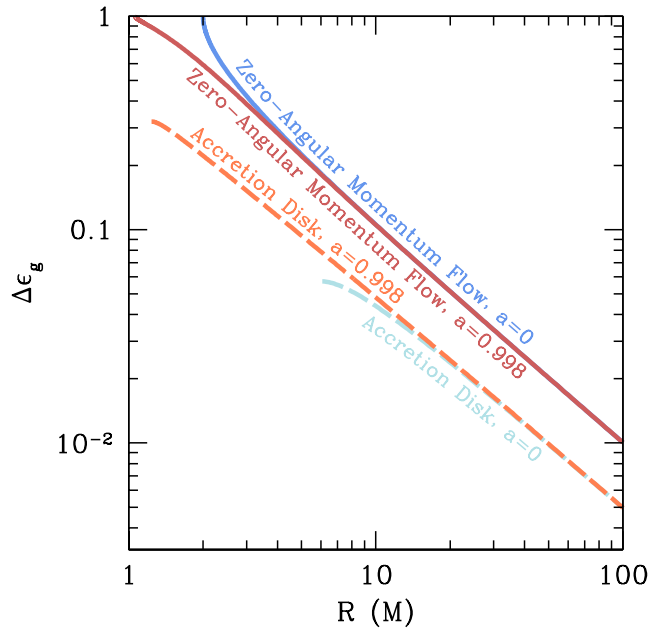}
\end{center}
\caption{{\em Left:} Binding energy released
  per unit rest mass as a function of radius for material that has
  come to rest in the zero-angular momentum frame ({\em solid lines})
  and the Keplerian frame ({\em dashed lines}) around a GR black hole.
  This is shown for a non-rotating GR black hole (Schwarzschild) by
  the blue lines and for a rapidly-rotating $(a=0.998)$ GR black hole
  (Kerr) by the orange lines.  Note that in both cases, for the
  zero-angular momentum flow the entire rest mass is released at the
  horizon, while the energy released by an accretion disk peaks at the
  ISCO (inside of which no further stable orbits exist).  Finally, beyond
  approximately $10 M$, both follow their Newtonian expressions.
}\label{fig:be}
\end{figure}

The gravitational binding energy released by a particle falling onto
the putative surface depends upon the details of the gravitational
theory.  For those theories which admit notions of energy conservation
(in the test particle limit), including all stationary spacetimes, we
may write this in terms of the specific binding energy at the putative
surface, $\Delta\epsilon_g$.  That is, the liberated energy due to a
particle of mass $m$ as measured at infinity is $E_\infty =
\Delta\epsilon_g m c^2$.  For a continuous accretion flow, this
provides a total power, as measured at infinity, of
\begin{equation}
L_\infty = \Delta\epsilon_g \dot{M} c^2\,,
\end{equation}
where $\dot{M}$ is the mass accretion rate at the surface.

The particular form of $\Delta\epsilon_g$ also depends upon the nature
of the accretion flow.  For arbitrary stationary, spherically
symmetric spacetimes this is computed in Appendix \ref{sec:GBEiSS} for
two typical cases:  zero-angular momentum accretion flows in which
matter does not orbit the black hole (e.g., Bondi accretion) and
Keplerian accretion disk (e.g., appropriate for thin disks).  In the
latter case $\Delta\epsilon_g$ is generally smaller since a fraction
of the liberated binding energy is necessarily converted into the
kinetic energy associated with the orbital motion.  Within the context
of GR black holes\footnote{Througout this section   we
  will punctuate the analysis   with examples from GR black holes.
  However, it should be   understood that our analysis is independent
  of   these specific   examples, and in fact considerably more
  general.} $\Delta\epsilon_g$ is straightforward to compute
explicitly for these cases, and is shown in Fig. \ref{fig:be} for both
nonrotating and rapidly-rotating black holes.  However, we shall see
that our ultimate constraints are independent of the form of
$\Delta\epsilon_g$.

Only a fraction of $L_\infty$ is converted into radiation.  The
observed electromagnetic luminosity at infinity may be parametrized in
terms of a radiative efficiency, $\eta_r$:
\begin{equation}
L_{\rm obs} = \eta_r L_{\infty}\,.
\end{equation}
Gravitational binding energy may also be converted into the kinetic
energy of relativistic outflows, which we similarly parametrize in
terms of outflow efficiency, $\eta_k$:
\begin{equation}
L_{\rm out} = \eta_k L_{\infty}\,.
\end{equation}
It is important to note that $\eta_r$ and $\eta_k$ are primarily a
function of the accretion flow, dependent upon the microphysics, and
thus relatively insensitive to the character of strong gravity.

If a horizon is present, the remainder of $L_\infty$ may be advected
across it without further observational consequence.  However, in the
presence of a surface, if Sgr A* has reached steady state, this
remainder must ultimately be radiated.  Thus the surface luminosity,
as measured at infinity, is
\begin{equation}
L_{\rm surf}
=
L_\infty - L_{\rm obs}-L_{\rm out}
=
\frac{1-\eta_r-\eta_k}{\eta_r} L_{\rm obs}\,,
\label{eq:Ls}
\end{equation}
Where we have writen this entirely in terms of the unknown
efficiencies and the observed luminosity.  (Specifically, note that
neither $\Delta\epsilon_g$ nor $\dot{M}$ appear in this expression.)

As we have argued in \S\ref{sec:PA.COaB}, for compact surfaces this
radiation will be in the form of a blackbody spectrum.  That is,
in terms of the apparent radius ($R_a$) and temperature ($T_\infty$)
of the putative surface, as measured at infinity,
\begin{equation}
L_{\rm surf} = 4\pi \sigma R_a^2 T_\infty^4\,.
\end{equation}
Alternatively, this provides a means to estimate the surface
temperature (and thus spectrum) given $L_{\rm surf}$ and $R_a$.
Combined with eq. (\ref{eq:Ls}), we may write $T_\infty$ in terms of
$L_{\rm obs}$, $R_a$, $\eta_r$ and $\eta_k$:
\begin{equation}
T_\infty =
\left(
\frac{1-\eta_r-\eta_k}{\eta_k} \frac{L_{\rm obs}}{4\pi \sigma R_a^2}
\right)^{1/4}\,.
\end{equation}
In terms of this surface temperature, the expected flux seen by
distant observers is then
\begin{equation}
F_\nu = \pi \left(\frac{R_a}{D}\right)^2 B_\nu\left(T_\infty\right)\,,
\end{equation}
where $B_\nu$ is the blackbody spectrum.

However, no thermal surface component is observed in Sgr A*'s
spectrum.  Thus, if such surface emission is present, it must be
hidden under the emission from the accretion flow.  As a consequence, each
flux measurement constitutes an independent upper limit upon the
surface flux, and thus surface temperture.  Explicitly, given an
observed flux
$F^{\rm obs}_\nu$, $T_\infty \le T_{\rm max}\left(\nu,F^{\rm obs}_\nu,R_a/D\right)$
where 
\begin{equation}
T_{\rm max}\left(\nu,F^{\rm obs}_{\nu};\frac{R_a}{D}\right)
=
h\nu \bigg/ k \ln\left(
1
+
\frac{2 \pi h \nu^3 R_a^2}{c^2 F^{\rm obs}_{\nu} D^2}
\right)\,.
\end{equation}
Note that this is generally a function of $R_a$, a consequence of the
fact that larger surfaces are correspondingly cooler, and therefore
easier to hide under the observed emission.

\begin{figure}
\begin{center}
\includegraphics[width=\columnwidth]{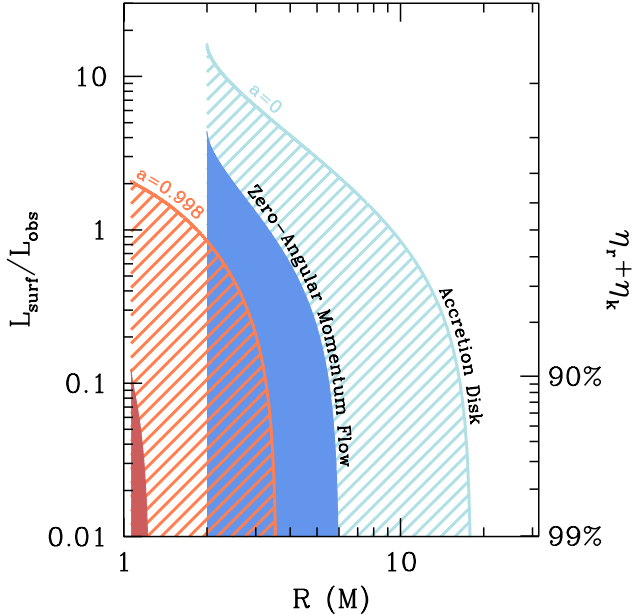}
\end{center}
\caption{Fundamental constraints upon $L_{\rm surf}/L_{\rm obs}$ (or
  $\eta_r+\eta_k$) in GR black hole spacetimes as a function of the
  coordinate radius of the stopping region.  In the shaded/hatched 
  regions
  $\eta_r+\eta_k > \Delta\epsilon_g(R)/\Delta\epsilon_g(r_{\rm ISCO})$
  and are thus not possible in the context of GR accretion theory.
  This is shown for when the accreted material comes to rest in the
  zero-angular momentum frame and the Keplerian frame around
  non-rotating and rapidly-rotating $(a=0.998)$ black holes.  In all
  cases, the shaded regions are truncated at the coordinate position
  of the relevant horizons and the excluded $L_{\rm surf}/L_{\rm obs}$
  vanishes at some radii, which is a natural result of fixing
  $r_{\rm ISCO}$ (for the zero-angular momentum flow, this happens at
  $r_{\rm ISCO}$ itself).
}\label{fig:etas}
\end{figure}

This maximum temperature then implies a limit upon $L_{\rm surf}$
directly via the black body condition:
\begin{equation}
\frac{L_{\rm surf}}{L_{\rm obs}}
\leq
\frac{L_{\rm surf, max}}{L_{\rm obs}}
\equiv
\frac{\sigma R_a^2}{D^2 F_{\rm obs}} T_{\rm max}^4\left(\nu,F^{\rm obs}_\nu;\frac{R_a}{D}\right)\,,
\label{eq:LsLalim}
\end{equation}
where $F_{\rm obs}=L_{\rm obs}/4\pi D^2$ is the integrated observed flux from
Sgr A*.  Alternatively, we may rewrite this in terms of a {\em lower}
limit upon the efficiencies:
\begin{equation}
\eta_r + \eta_k \ge \frac{1}{1 + L_{\rm surf, max}/L_{\rm acc}} \,.
\end{equation}
That is, if the efficiencies are sufficiently high the unradiated
liberated energy in the accretion flow is sufficiently small that the
surface could escape detection.  Unlike $L_{\rm surf}/L_{\rm acc}$, we
have natural scales against which to compare $\eta_r+\eta_k$.
In prodigiously accreting systems, such as active galactic
nuclei, X-ray binaries \& gamma-ray bursts, which are believed to be
radiatively efficient, the radiative efficiency $\eta_r+\eta_k$ is estimated
to be $\sim 10\%$ \citep{Kato:08}.  In Sgr A*, on the other hand, $\eta_r$ is
thought to be quite small ($0.01$--$1\%$ for typical accretion models)
as a consequence of the weak coupling between accreting ions and
electrons within the gas \citep{Nara-McCl:08}, and there is presently no
direct evidence for energetic outflows.

Within the context of GR black holes and standard accretion theory,
there are fundamental limits upon how large $\eta_r+\eta_k$ may be for
compact surfaces.  This arises from the fact that matter inside of the
Innermost Stable Circular Orbit (ISCO) plunges rapidly onto the
surface, and thus does not have sufficient time to radiate.  In
contrast, outside of the ISCO material may linger on stable orbits for
long periods of time, and at least in principle can radiate
efficiently.  Thus, even if the ``intrinsic'' radiative efficiency
outside of the ISCO reached unity, the accreting material would still
accrue additional kinetic energy during the plunge from the ISCO to
the putative surface.  If this surface is located inside of the ISCO,
we may therefore place a lower bound upon the additionally liberated
energy, and thus an upper bound upon $\eta_r+\eta_k$.  These are shown
in Fig. \ref{fig:etas} for zero-angular momentum accretion
flows\footnote{Note that we are 
being maximally conservative by assuming that a zero-angular momentum
flow can radiate all its binding energy down to the ISCO.  Models of
non-rotating accretion flows are usually very radiatively inefficient.
Allowing for this would strengthen our argument substantially.}  and
orbiting accretion disks surrounding both non-rotating and
rapidly-rotating black holes (see Appendix \ref{sec:GBEiSS} for
details).  For orbiting disks the maximum value of
$\eta_r+\eta_k$ is generally less than $33\%$.

Therefore, each combined size and flux measurement of Sgr A* places a
direct constraint upon the luminosity of a putative surface, or,
equivalently, upon the radiative efficiency of the accretion flow.
Thus, we now turn our attention to describing the current best
constraints upon $R_a/D$ and $F^{\rm obs}_\nu$.

\subsection{Millimeter Size Constraints}\label{sec:OLUtEoH.MSC}
Measuring the intrinsic size of Sgr A* is complicated by its low
luminosity, exceedingly small size, interstellar scattering and the
opacity of the surrounding accretion flow.  As a result,
high-frequency VLBI has produced the only meaningful limits upon the
size of a putative surface.  Recent measurements of the photosphere
radius (half-width, half-max) are listed in Table \ref{tab:VLBI}.

\begin{deluxetable}{ccccc}
\tablecaption{VLBI Size Constraints\label{tab:VLBI}}
\tablehead{
\colhead{$\lambda$ ($\mm$)} &
\colhead{$R_a$ ($\muas$)\tablenotemark{a}} &
\colhead{$3$--$\sigma R_a$ ($\muas$)\tablenotemark{b}} &
\colhead{Ref.}
}
\startdata
$1.3$ & $19$  & $27$  & \citet{Doel_etal:08}\\
$3$   & $63$  & $84$  & \citet{Shen_etal:05}\\
$7$   & $120$ & $150$ & \citet{Bowe_etal:04}
\enddata
\tablenotetext{a}{The measured half-width, half-max}
\tablenotetext{b}{The $3$--$\sigma$ upper limit upon the half-width, half-max}
\end{deluxetable}

At wavelengths longer than $1.3\,\mm$ Sgr A* is optically thick and
the observed size is dominated by interstellar scattering, requiring a
careful subtraction of the empirically determined scattering law.
However, at $1.3\,\mm$ interstellar scatting is subdominant and the
plasma surrounding Sgr A* has become optically thin.  Perhaps
surprisingly, at this wavelength the inferred angular size,
$37\pm11\,\muas$, is {\em smaller} than the apparent diameter of the
horizon ($48$--$55\,\muas$ depending upon spin), {\em independent of
  black hole spin} \citep{Doel_etal:08}.  However, this is natural in
the context of an orbiting accretion flow or an outflow; in both cases
the image is dominated by the approaching
plasma \citep{Brod-Fish-Doel-Loeb:08,Brod-Loeb:09}.  Therefore, the
measured $37\pm11\,\muas$ may not represent $R_a/D$ of the surface
itself, but instead result from the velocity structure of the
accreting material. Nevertheless, efforts to fit this result using
existing accretion models for Sgr A* imply that the surface lies
within the photon orbit.

\subsection{Infrared Flux Limits}\label{sec:OLUtEoH.IFL}
Given the size limits, typical values of $T_\infty$ range from
$10^2$--$10^4\,\K$, and thus the emission from a putative surface
peaks in the near and mid-infrared.  Fortunately, despite the large
infrared extinction in the direction of the Galactic center, there are
a number of observations of Sgr A* at these wavelengths.  Furthermore,
there exists an accurate empirically determined extinction
law for this region \citep{Mone_etal:01}, making it possible to
deredden the observed fluxes, producing intrinsic infrared flux
measurements for Sgr A*.

\begin{deluxetable}{cccc}
\tablecaption{Dereddened Infrared Flux Limits\label{tab:IR}}
\tablehead{
\colhead{Band} &
\colhead{$\lambda$ ($\mum$)} &
\colhead{$F_\nu$ ($\rm mJy$)\tablenotemark{a}} &
\colhead{Ref.}
}
\startdata
$H$   & $1.6$ & $2.60\pm0.28$ & \cite{Horn_etal:07}\\
$K'$  & $2.1$ & $1.66\pm0.18$ & \cite{Horn_etal:07}\\
$L'$  & $3.8$ & $1.28\pm0.30$ & \cite{Ghez_etal:05b}\\
$M_S$ & $4.7$ & $2.2\pm1.6$   & \cite{Horn_etal:07}\\
$N$   & $8.7$ & $22\pm14$     & \cite{Scho_etal:07}
\enddata
\tablenotetext{a}{The uncertainty in $F_\nu$ is taken to be the
  $1-\sigma$ detection uncertainty.}
\end{deluxetable}

However, in these bands the emission from Sgr A* is dominated by
variability \citep[see, e.g.,][]{Horn_etal:07}, which must clearly be
associated with the accretion flow and not the surface.  This prevents
an unambiguous detection of quiescent emission.  Therefore, the flux
limits we collect in Table \ref{tab:IR} are determined by inspecting
variable infrared light curves and extracting the lowest flux
detection.  The $L'$ ($3.8\,\mum$) and $N$ ($8.7\,\mum$) band limits
are taken directly from \cite{Ghez_etal:05b} and \cite{Scho_etal:07},
respectively.  The $H$ ($1.6\,\mum$), $K'$ ($2.1\,\mum$) and $M_S$
($4.8\,\mum$) band limits were determined by examining the panels of
Fig. 2 of \citet{Horn_etal:07}, which show dereddened flux light
curves for a number of infrared flares in Sgr A* and a variety of
near-infrared bands.

\subsection{Constraints Upon Surface Existence}\label{sec:OLUtEoH.CUSE}

\begin{figure}
\begin{center}
\includegraphics[width=\columnwidth]{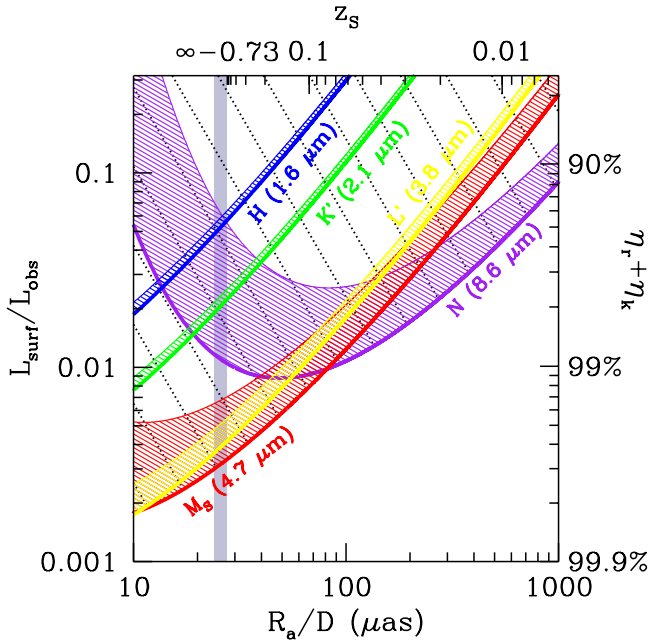}
\end{center}
\caption{The limits upon $L_{\rm surf}/L_{\rm obs}$
  implied by eq. (\ref{eq:LsLalim}) as a function of the photosphere
  size as seen at infinity for the infrared measurements listed in
  Table \ref{tab:IR}.  The hatched bands denote the $3$--$\sigma$ upper-bounds.
  The peculiar behavior of the $N$-band constraint is a result of the
  transition from the Rayleigh-Jeans limit to the Wien limit around
  $R_a/D\simeq50\,\muas$ as the surface becomes cooler.  The region
  above any of the limits is necessarily excluded.  The right-hand
  vertical axis shows the corresponding limits upon the accretion
  flow's radiative efficiencies.  The top axis gives the redshift
  associated with a Schwarzschild spacetime given the apparent source
  radius and the thick grey line shows the apparent radii associated
  with the photon orbit for Kerr spacetimes.}\label{fig:irlim}
\end{figure}

\begin{figure}
\begin{center}
\includegraphics[width=\columnwidth]{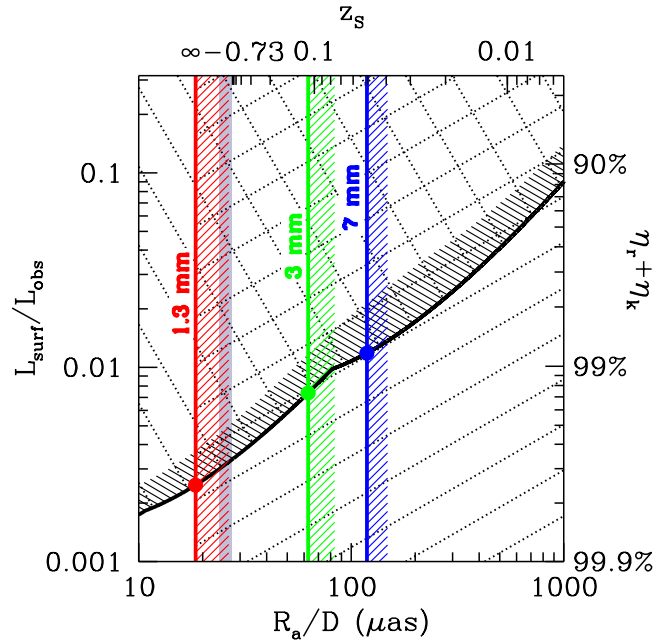}
\end{center}
\caption{The limits upon $R_a/D$ implied by recent VLBI observations
  listed in Table \ref{tab:VLBI}, overlayed upon the combined constriant
  implied by IR flux measurements.  Regions to the right of the
  left-most (smallest) size constraint are excluded.  When combined
  with the limits from the IR flux measurements, the permissible
  parameter space is reduced to a small corner in the
  $R_a$--$L_{\rm surf}/L_{\rm obs}$ plane.  The axes are identical to
  those in Fig. \ref{fig:irlim}.}\label{fig:mmlim}
\end{figure}

Each of the infrared flux limits listed in Table \ref{tab:IR} places
an upper limit upon $L_{\rm surf}/L_{\rm obs}$ via
eq. (\ref{eq:LsLalim}), given a value of $R_a$.  Fig. \ref{fig:irlim}
shows these limits as a function of surface size, together with their
$3$--$\sigma$ upper bounds (denoted by the hatched regions).
Generally, larger surfaces are cooler.  This has two consequences.
First, since the strongest constraints are placed by flux measurements
near the peak in the thermal spectrum, the infrared band providing the
most stringent limit is a function of $R_a$ as well; near
$R_a/D\simeq100\,\muas$ the infrared band dominating the constraint
changes from $M_S$ ($4.7\,\mum$)\footnote{Though the small
  uncertainties associated with the $L'$ ($3.8\,\mum$) measurement
  dominate the uncertainties at low $R_a$ despite the fact that the
  limit derived from the best estimate of the $M_S$-band flux is
  always stronger.} to $N$ ($3.8\,\mum$).  The constraint due to
the combination of all of the infrared bands is the lower-envelope of
all the bands, and defines the excluded region, corresponding to
small, luminous surfaces.  Second, since cooler surfaces are more
easily masked by the emission from the accretion flow, the limit
becomes less stringent as $R_a$ increases.  Thus, the infrared
constraint upon $L_{\rm surf}/L_{\rm obs}$ must be supplemented with
an independent limit upon the size of the putative surface.

Upper limits upon $R_a/D$ are found directly via the radio VLBI
observations collected in Table \ref{tab:VLBI}.  These are shown in
Fig. \ref{fig:mmlim}, with their $3$--$\sigma$ upper bounds (again
denoted by the hatched regions) together with the combined infrared
limit.  The recent $1.3\,\mm$ detection is the strongest, and excludes
$R_a/D>27\,\muas$ at the $3$--$\sigma$ level.  When combined with the
infrared constraints, only a small corner of the
$R_a$--$L_{\rm surf}/L_{\rm obs}$ is still permitted, corresponding to
small, dim surfaces.  Earlier measurements at $7\,\mm$ and
$3\,\mm$ already required $L_{\rm surf}/L_{\rm obs} \lesssim 0.02$.
The new $1.3\,\mm$ detection improves these by nearly an order of
magnitude, requiring $L_{\rm surf}/l_{\rm obs} \lesssim 0.004$ at the
$3$--$\sigma$ level.

A more physical interpretation is provided by the accretion
efficiencies the latest observations now demand, shown on the
right-hand vertical axis.  In order for a surface to be present
$\eta_r+\eta_k \gtrsim 99.6\%$ (at $3$--$\sigma$)!  That is,
as matter falls onto Sgr A*, somehow $99.6\%$ of the liberated
gravitational binding energy must be radiated, powering either the
observed luminosity or kinetic outflows.  Otherwise the emission of
the remainder upon settling onto the surface would have been detected.
This needed efficiency is considerably larger than even the $10\%$
efficiencies implicated in rapidly accreting systems, let alone the
meager efficiencies ($0.01-1\%$) inferred in Sgr A*.

\begin{figure}
\begin{center}
\includegraphics[width=\columnwidth]{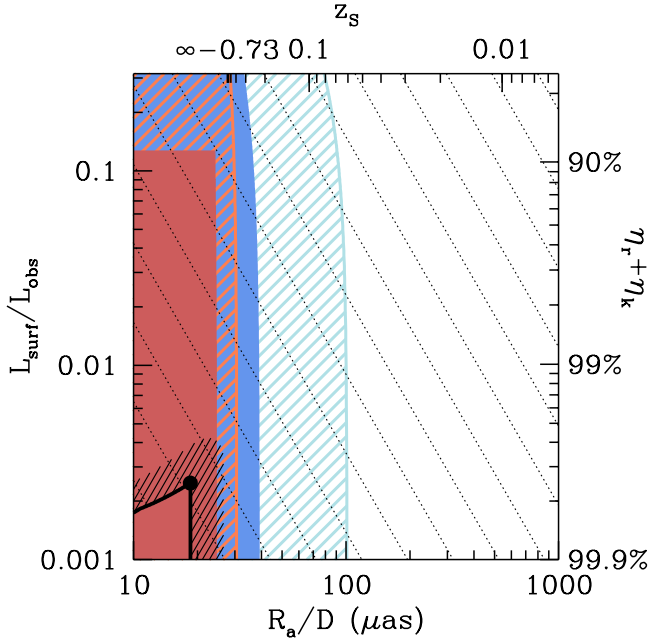}
\end{center}
\caption{The combined limits upon $L_{\rm surf}/L_{\rm obs}$ and
  $R_a/D$ compared with the regions generally excluded in GR black
  hole spacetimes (shown as a function of coordinate radius in
  Fig. \ref{fig:etas}).  Extending these to angular scales smaller
  than the theoretical minimum (the angular size of the horizon)
  results in the rectangular appearance.  Blue and red
  excluded regions correspond to efficiencies above the maximum
  possible values for accretion onto Schwarzschild and maximal-Kerr
  black holes, respectively.  Hatched and solid limits correspond to
  Keplerian and zero-angular momentum accretion flows, respectively.
  Given the most recent infrared flux limits and mm-VLBI size constraints,
  $\eta_r+\eta_k$ is inconsistent by more than an order of magnitude
  with estimates for Sgr A* specifically, relativistic accretion flows
  generally and even fundamental limits in GR spacetimes.  The axes
  are identical to those in Fig. \ref{fig:irlim}.}\label{fig:grlim}
\end{figure}

More striking is the comparison of this to the {\em maximum}
efficiencies within GR spacetimes.  A variety of
black hole alternatives are grafted onto such spacetimes outside of
some compact region, and thus in these the GR limits
explicitly hold.  For other black hole alternatives, including those
associated with different gravity theories, these limits at least
provide a scale for comparison.  These fundamental constraints upon
the efficiencies arise from the fact that accreting material cannot
radiate efficiently inside of the innermost stable circular orbit
(ISCO).  Inside of the ISCO accreting material plunges inward rapidly
in comparison to the radiative timescales.  Thus, {\em even if the
  intrinsic radiative   efficiency outside the ISCO is 100\%} (which
is extremely unlikely), the fraction of the total binding energy
released by accreting gas in the course of its inward flow is limited
to that fraction that is emitted outside the ISCO.  
Therefore, within the context of GR, even with the most conservative of
assumptions (zero-angular momentum accretion flow with 100\% radiative
efficiency down to the ISCO) we obtain an absolute upper limit on
$\eta_r+\eta_k$ for surfaces which lie within the photon orbit of 91\%,
while for a more reasonable accretion scenario involving orbiting gas
the limit is $33\%$ (see Appendix \ref{sec:GBEiSS} for more details).
The combined observational limits upon $L_{\rm surf}/L_{\rm acc}$ and
$\eta_r+\eta_k$ are compared against those inferred in GR spacetimes
for zero-angular momentum and Keplerian flows in
Fig. \ref{fig:grlim}.  In particular, there is no longer any
allowed region given the recent $1.3\,\mm$ size constraint.

\section{Conclusions}\label{sec:C}

Recent infrared and mm-VLBI observations imply that if the matter
accreting onto Sgr A* comes to rest in a region visible to distant
observers, the luminosity associated with the surface emission from
this region satisfies $L_{\rm surf}/L_{\rm acc} \lesssim 0.003$.
Equivalently, these observations require that $99.6\%$ of the
gravitational binding energy liberated during infall is radiated in
some form prior to finally settling.  These numbers are inconsistent
by orders of magnitude with our present understanding of the radiative
properties of Sgr A*'s accretion flow specifically and relativistic
accretion flows generally.  Therefore, it is all but certain that no
such surface can be present, i.e., {\em an event horizon must exist}.

Critical to our analysis is the recent sub-mm constraint upon the size of
the emitting region of Sgr A* since it: (1) justifies the assertion
that the putative photosphere is sufficiently compact to be treated as a
blackbody, (2) limits the photosphere's temperature from below,
and (3) within the context of GR provides a
strong constraint upon possible values of the radiative efficiency
that is easily excluded.  Future mm-VLBI observations will be critical
to understanding the morphology of Sgr A*'s emitting region, and thus
validating the interpretation of current observations as the
approaching side of an orbiting accretion flow, with the attendant
implications for Sgr A*'s size.  However, unless the spacetime around
Sgr A* deviates substantially from that of a GR black hole, future
observations of the intrinsic size of Sgr A* will be unable to further
restrict the size of a putative surface due to the influence of
gravitational lensing.  The reason is that all objects which lie
within the photon orbit generally have the same apparent radius.
Thus, obtaining more sensitive near-infrared flux measurements 
will remain a crucial avenue for strengthening this kind of argument.

We were able to place these constraints completely in terms of
observable quantities: fluxes and the apparent size of Sgr A*.  Beyond
using GR spacetimes to provide scales for the purpose of comparison,
we did not need to make use of the particular structure of GR black
hole spacetimes.  As a result, our conclusions may be applied more
generally to all gravitational theories that admit notions of energy
conservation in the test-particle limit.  Specifically, these include
all geometric gravitational theories that admit stationary solutions,
including all of the $f(R)$ theories and black hole alternatives that
exist within the context of GR.  As a consequence, we cannot yet say
that Sgr A* is described by a GR black hole despite being able to
conclude that a horizon must exist.

\appendix

\section{Gravitational Binding Energy in Stationary Spacetimes}\label{sec:GBEiSS}

As material falls into the the gravitational potential well induced by
Sgr A* it necessarily converts a portion of its gravitational binding
energy into luminosity.  In \S\ref{sec:OLUtEoH} we simply parametrized
the amount of liberated energy in terms of some $\Delta\epsilon_g$,
finding it unnecessary to specify this function further.
In this appendix we derive a general
expression for the magnitude of the binding energy available.
We necessarily assume that gravity is described by a metric theory,
and that this theory admits a stationary solution.  For simplicity,
we will not explicitly discuss non-spherical solutions, however this
makes no formal difference.  We take the metric signature to be
$\,-+++$ and choose units such that $G=c=1$.

The state of an infalling particle, characterized by
its 4-momentum, $p^\mu$, can be used to determine the energy available to
radiation as measured at infinity.  Given an initial momentum for an
infalling particle, $p_i^\mu$, the momentum of the radiated photon,
$p_\gamma^\mu$, and the final particle momentum, $p_f^\mu$,
conservation of momentum gives
\begin{equation}
p_\gamma^\mu = p_i^\mu - p_f^\mu\,.
\end{equation}
The steady state of the spacetime implies that along geodesics (e.g., the
null geodesic followed by the photon, or the free-fall of the gas particle)
$p_t$ is explicitly conserved, corresponding to a conserved energy.  Thus,
the photon's energy at infinity is simply $\Delta \epsilon_g \equiv
-{p_\gamma}_t = {p_f}_t - {p_i}_t$.  Since this quantity is observed at
infinity, where the spacetime is flat, we are guaranteed that this quantity
is not affected by the choice of coordinates deep in the gravitational
well.  In addition, this has the virtue of being precisely what is
measured.

We will consider two scenarios, corresponding to different choices of
$p_f^\mu$.  The first of these is the most extreme: the particle was
initially at rest at infinity (${p_i}_t = -m$), fell to some radius
and came to rest, with \footnote{This
  corresponds to the momentum of an observer that has freely fallen
  from infinity and decelerated only along the direction of motion.
  In the context of general relativity this choice of rest frame
  corresponds to the Zero Angular Momentum Observer for rotating black
  holes; however, this expression is completely general.}
\begin{equation}
p^{\rm rest}_\mu = \left(\frac{m}{\sqrt{-g^{tt}}},0,0,0 \right)\,,
\end{equation}
at which point its accumulated binding energy was emitted.  The
available energy, per unit mass is then
\begin{equation}
\Delta \epsilon_g
=
\frac{p_t - p^{\rm rest}_t}{m}
=
\frac{\sqrt{-g^{tt}}-1}{\sqrt{-g^{tt}}}
=
\frac{z}{z+1}\,,
\end{equation}
where the redshift is defined in the usual way:
$z\equiv \sqrt{-g^{tt}}-1$.

Our second scenario involves the presence of an accretion disk.
Previously, we assumed that the accreting material has no angular
momentum, and therefore explicitly ignored orbital motion.  Generally,
the gas is expected to orbit the black hole, and
the kinetic energy in the orbital motion will be supplied by the
liberated gravitational binding energy, decreasing the reservoir of
energy available to radiation.  Hence we have placed a firm upper limit
upon the luminosity of the accretion flow itself.  However, we can
compute the decrease explicitly for the simple case of 
stationary, spherically symmetric spacetimes.  If
\begin{equation}
ds^2 = g_{tt} \d t^2 + g_{rr} \d r^2 + r^2 \d \Omega^2\,,
\end{equation}
for orbits in the equatorial plane, $p^{\rm orbit}_t$ and $p^{\rm orbit}_\phi$ are conserved,
$p_{\rm orbit}^\theta$ vanishes due to symmetry and,
\begin{equation}
p_{\rm orbit}^r = m \frac{\d r}{\d\tau}
=
\sqrt{-\frac{1}{g_{rr}}
\left( m^2 + g^{tt} {p^{\rm orbit}_t}^2 + \frac{{p^{\rm orbit}_\phi}^2}{r^2} \right)} = 0\,.
\end{equation}
If the orbit is to remain closed, we also require
\begin{equation}
m \frac{\d^2 r}{\d\tau^2}
=
-\frac{1}{2} \frac{\partial}{\partial r} 
\frac{1}{g_{rr}}
\left( m^2 + g^{tt} {p^{\rm orbit}_t}^2 + \frac{{p^{\rm orbit}_\phi}^2}{r^2} \right)
=
0\,.
\end{equation}
Solving these for $p^{\rm orbit}_t$ and $p^{\rm orbit}_\phi$ gives
\begin{equation}
p^{\rm orbit}_t = \frac{1}{\sqrt{-g^{tt}}} \bigg/
\sqrt{1 - \frac{1}{2} \frac{\partial \ln(-g^{tt})}{\partial \ln r }}
=
\frac{1}{z+1} \bigg/
\sqrt{ 1 + \frac{\partial \ln(z+1)}{\partial \ln r} }\,,
\end{equation}
which is generally larger than for the zero-angular momentum case (since
$z$ generally decreases with radius).  Correspondingly, $\Delta\epsilon_g$
in this case is generally lower.

While our discussion of the liberated gravitational binding energy is
quite general, requiring only that gravity is described by a metric
theory that admits a stationary solution, GR can provide some
intuition regarding the magnitude of energy that can reasonably be
released.  Fig. \ref{fig:be} shows the specific binding energy released
by material as a function of radius.  Typically, the ISCO is the final
radius at which matter can efficiently radiate since inside this point
material plunges into the black hole on a free fall
timescale
\footnote{In the context of thin disks this
  corresponds to the zero-torque inner-boundary condition, which has
  been well tested by the fitting of X-ray spectra from X-ray
  binaries.  In the context of thick disks, or quasi-spherical
  accretion flows, in principle magnetic fields can couple material
  inside of the ISCO to the flow outside, transferring energy in the
  process.  However, thick disks are thick {\em because}
  $\eta_r\lesssim 0.1$, limiting the ability of the disk to cool
  efficiently.  Thus, even in the presence of substantial magnetic
  coupling, thick disks are also bounded by this limit.
},
rapid in comparison to the relevant radiative timescale for the
rather pedestrian densities and magnetic fields observed in Sgr A*
\footnote{In order to reach $\eta_r+\eta_k\simeq1$ the energy of the
  accreting material must be extracted over the timescale comparable
  to the free-fall timescale at the point where $\eta_r+\eta_k$ of the
  binding energy has been released.  That is, because the liberated
  binding energy rapidly increases as material falls inward, $\eta_r+\eta_k$
  is determined by the last radius at which energy extraction can keep
  pace with the infall.  Explicitly, for a Schwarzschild black hole
  this implies that the cooling timescale must be less than
  $(1-\eta_r-\eta_k) \Delta\epsilon_g/(\d\Delta\epsilon_g/\d\tau) = 2 R (R/2M)^{3/2} \sqrt{1-2M/R} (1-\eta_r-\eta_k)$.
  For $\eta_r+\eta_k=99.6\%$, this corresponds to cooling timescales of
  $2\,\s$ at $R=6M$ and $0.3\,s$ at $R=3M$, both of which are only
  realized in practice in extraordinarily dense environments such as
  newly formed neutron stars.}.
As a consequence, {\em even
  if 100\% of the gravitational binding energy released outside of the
  ISCO is radiated or goes into outflows}, the maximum fraction of the rest mass that can be
emitted is shown by the blue line in Fig. \ref{fig:be}.  If the region
where the accreted material finally comes to rest lies within the
ISCO, additional gravitational binding energy will be released, and
must subsequently be radiated in the stopping region.  Thus, the
radiative efficiency of the accretion flow, defined by
\begin{equation}
\eta_r + \eta_k \equiv
\frac{L_{\rm acc}}{\Delta\epsilon_g(R) \dot{M}}
+
\frac{L_{\rm out}}{\Delta\epsilon_g(R) \dot{M}}
\end{equation}
must satisfy
\begin{equation}
\eta_r+\eta_k \le \frac{\Delta\epsilon_g(r_{\rm ISCO})}{\Delta\epsilon_g(R)}
\end{equation}
where $R$ is the radius of the stopping region.  This limit shown as a
function of $R$ in Fig. \ref{fig:etas} for zero-angular momentum
accretion flows and accretion disks in Schwarzschild and Kerr
spacetimes.  Generally, once $R$ is constrained to lie within the
photon orbit, the maximum $\eta_r+\eta_k$ possible is $91\%$, corresponding
to a zero-angular momentum accretion flow in a rapidly-rotating
($a=0.998$) Kerr spacetime.  When an accretion disk is present this
limit declines to $33\%$.  For the Schwarzschild spacetime, $\eta_r+\eta_k$ is
bounded from above by $43\%$ (zero-angular momentum accretion) and
$14\%$ (accretion disk).  In all cases, these maximum values are
substantially smaller than the limits set by recent sub-mm and infrared
observations if Sgr A* did not have a horizon.

\section{Apparent Sizes of Compact Objects}\label{sec:ASoCO}

 Observations of the size of a compact
emitting region is necessarily impacted by strong gravitational
lensing.  In metric theories of gravity, objects associated
with deep potential wells will appear larger to observers at infinity.
The apparent size of the region is directly related to both the
physical size and the redshift of the compact object.  Thus, relating
the actual size of a compact emitting region to the observed size
requires some understanding of the spacetime structure around the
object.  This is, of course, one of the reasons we chose to cast the
constraints upon the existence of a horizon, described in
\S\ref{sec:OLUtEoH}, in terms of $R_a$ and not a physical object
size.  Nevertheless, for completeness, we discuss the procedure here.

It is typically very difficult to compute the relationship between the
physical and observed size and shape of a compact emitting object.
However, in the special case of a spherically symmetric spacetime, this is
generally tractable, independent of the particular form of the
metric.  We assume
\begin{equation}
ds^2 = g_{tt} \d t^2 + g_{rr} \d r^2 + r^2 \d \Omega^2\,,
\end{equation}
where $g_{tt}$ and $g_{rr}$ are functions of $r$ alone.  Then, in the
equatorial plane, the null geodesics are defined by
\begin{equation}
\frac{\d t}{\d\lambda} = g^{tt}
\,,\quad
\frac{\d r}{\d\lambda} = \sqrt{
\frac{1}{g_{rr}} \left(
-g^{tt}-\frac{b^2}{r^2}
\right)}
=
\sqrt{\frac{-g^{tt}}{g_{rr}}\left[ 1 - \frac{b^2}{(z+1)^2r^2} \right] }
\,,\quad
\frac{\d\theta}{\d\lambda} = 0
\quad{\rm and}\quad
\frac{\d \phi}{\d\lambda} = \frac{b}{r^2}\,,
\end{equation}
where the equations for $\d t/\d\lambda$ and $\d\phi/\d\lambda$ are
associated with the existence of a time-like and azimuthal Killing
vectors, respectively, the equation for $\d\theta/\d\lambda$ is fixed
by vertical symmetry and the equation for $\d r/\d\lambda$ arises from
the null ray condition.  In these $b$ is the impact parameter at
infinity and $\lambda$ is an arbitrary affine parameter.  The minimum
radius reached by a given null geodesic occurs at its inner turning
point, at which $ b = r \sqrt{-g^{tt}} = r ( z + 1 )$.  Alternatively,
this corresponds to the maximum $b$ that a null geodesic can have and
still impact a surface of radius $r$.  Thus, the apparent radius,
$R_a$, of an object with physical radius $R$ is simply
\begin{equation}
R_a = R \left( z + 1\right)\,.
\end{equation}

Some care must be taken, however, when $R$ is smaller than the photon
orbit.  This is because rays which cross the photon orbit have no radial
turning points, and therefore will in all cases be captured.  This is
clear from the definition of the photon orbit, $r_\gamma$:
\begin{equation}
\left. \frac{\d}{\d r} \frac{1}{(z+1)^2 r^2} \right|_{r_\gamma}= 0\,,
\end{equation}
which corresponds to the position of the maximum of the ``effective
potential'' in the radial equation.  Thus, if
\begin{equation}
b < r_\gamma \bigg[z(r_\gamma)+1\bigg]
\qquad\Rightarrow\qquad
\frac{b^2}{(z+1)^2 r^2} < 1
\quad\text{for all}\quad r\,.
\end{equation}
As a consequence, the apparent radius of objects for
which $R<r_\gamma$ is the same as that for objects with $R=r_\gamma$.
Thus, generally,
\begin{equation}
R_a = \left\{
\begin{aligned}
&r_\gamma \bigg[ z(r_\gamma) + 1 \bigg] &&\text{if }R\le r_\gamma\\
&R \bigg[ z(R) + 1 \bigg] &&\text{otherwise}\,.
\end{aligned}
\right.
\end{equation}

In the case of the Schwarzschild metric, this gives the well-known result
\begin{equation}
R_a = \left\{
\begin{aligned}
&3 \sqrt{3} M && \text{if } R\le 3M\\
&R \sqrt{\frac{R}{R-2M}} && \text{otherwise}\,.
\end{aligned}
\right.
\end{equation}
For a rapidly rotating Kerr spacetime, the apparent radius in the
equatorial plane may also be computed without undue difficulty (though
in this case care must be taken into account for the non-diagonal
components of the metric).  Generally, this is given by
\begin{equation}
R_a = \frac{1}{2}\left(b_+ - b_-\right)\,,
\end{equation}
where
\begin{equation}
\begin{aligned}
b_\pm &= \pm {\rm max}\left(
R \frac{R\sqrt{R^2-2MR+a^2} \mp 2 a M}{R^2-2MR}
\,,
r_{\pm\gamma}
\frac{r_{\pm\gamma}\sqrt{r_{\pm\gamma}^2-2Mr_{\pm\gamma}+a^2} \mp 2 a M}
{r_{\pm\gamma}^2-2M r_{\pm\gamma}} \right)
\end{aligned}
\end{equation}
in which $r_{\pm\gamma}$ is the radius of the prograde/retrograde
photon orbit.  Since these differ for rotating black holes, we
generally have three conditions.  In the case of a maximally rotating
black hole ($a=1$), this expression is especially simple:
\begin{equation}
R_a = \left\{
\begin{aligned}
& \frac{9}{2} M && \text{if } R\le r_{+\gamma}\\
& \frac{R + 8M}{2} && \text{if } r_{+\gamma} < R \le 4M\\
& R \frac{R-1}{R-2} && \text{otherwise}\,,
\end{aligned}
\right.
\end{equation}
where $R$ is the object radius in Boyer-Lindquist coordinates.  While
$r_{+\gamma}=M$ in these coordinates for $a=1$, there remains a finite
proper distance between the photon orbit and the horizon, the equality
being an artifact of the coordinates themselves.  For this reason, we
distinguish between these formally, though it makes no difference
(since $R_a=9/2$ for all $R$ between $r_{+\gamma}$ and the horizon),
as it must not given that $R_a$ is a gauge invariant quantity as a
consequence of its definition.

Most important for the present discussion is the fact that $R_a$ for the
Schwarzschild and equatorial Kerr spacetimes differs by only about $15\%$.
Thus, despite the vastly different coordinate sizes, an object with $R=1M$
embedded in a maximal Kerr spacetime has roughly the same apparent size
as an object with $R=3M$ embedded in a Schwarzschild spacetime.  
As a consequence, if the present limit upon the size of the sub-mm emitting
region in Sgr A* of $37\,\muas$ ($R_a=3.5 M$) is interpreted as a
photosphere surrounding the stopping region, it constrains the size of a
central emitting region to lie well within the photon orbit of both a Kerr
and Schwarzschild black hole.

On the other hand, the anomalously small apparent radius might appear
unphysical within the context of GR.  However, this is easily
rectified if the emission region is interpreted instead as the visible
arc of an oncoming accretion disk \citep[as a result of Doppler
boosting and Doppler shifts, the receding side being considerably
dimmer for the same reason,][]{Brod-Loeb:06}.  While the equatorial
extent of the arc can be significantly smaller than the minimum
apparent radius in this situation, the vertical extent is still roughly
$2R_a$ \citep[see, e.g.,]{Brod-Fish-Doel-Loeb:08,Brod-Nara:06}.  Hence,
unless the projected baseline was extraordinarily fortuitously
aligned, again we would expect large measured sizes for central
emission regions larger than the photon orbit radius.  It is possible
to coincidentally fit the existing spectral, polarization and mm-VLBI
observations using orbiting accretion flow models
\citep{Brod-Fish-Doel-Loeb:08}.  However, this is due at least in part
to the fact that the existing mm-VLBI size constraint is essentially
restricted to the East-West direction.  Future mm-VLBI observations
will be critical to unambiguously determining the morphology of the
emitting region \citep{Fish-Brod-Doel-Loeb:09}.

\end{document}